\documentclass{icrc2009}

\usepackage{graphicx}   
\usepackage{caption}    
\usepackage[font=footnotesize]{subfig} 
\usepackage{fixltx2e}
\usepackage{url}

\newcommand{\shorttitle}[1]%
{\markboth{Proceedings of the 31\MakeLowercase{$^{st}$} ICRC, {\L}\'{o}d\'{z} 2009}{#1} }
\newcommand{\etal}{\MakeLowercase{\textit{et al. }}} 

\usepackage{epstopdf}

\begin{document}
\title{The readout system of the MAGIC-II Cherenkov Telescope}

\author{\IEEEauthorblockN{
D. Tescaro\IEEEauthorrefmark{1}, 
J. Aleksic\IEEEauthorrefmark{1}, 
M. Barcelo\IEEEauthorrefmark{1}, 
M. Bitossi\IEEEauthorrefmark{2}, 
J. Cortina\IEEEauthorrefmark{1}, 
M. Fras\IEEEauthorrefmark{3},
D. Hadasch\IEEEauthorrefmark{4},
J. M. Illa\IEEEauthorrefmark{1}, \\
M. Mart\'{\i}nez\IEEEauthorrefmark{1}, 
D. Mazin\IEEEauthorrefmark{1}, 
R. Paoletti\IEEEauthorrefmark{5}, 
R. Pegna\IEEEauthorrefmark{5}
on behalf of the MAGIC collaboration}
\\
\IEEEauthorblockA{\IEEEauthorrefmark{1}IFAE, Edifici Cn., Campus UAB, E-08193 Bellaterra, Spain}
\IEEEauthorblockA{\IEEEauthorrefmark{2}Universit\`a di Pisa, and INFN Pisa, I-56126 Pisa, Italy}
\IEEEauthorblockA{\IEEEauthorrefmark{3}Max-Planck-Institut f\"ur Physik, D-80805 M\"unchen, Germany}
\IEEEauthorblockA{\IEEEauthorrefmark{4}Universitat Aut\`onoma de Barcelona, E-08193 Bellaterra, Spain}
\IEEEauthorblockA{\IEEEauthorrefmark{5}Universit\`a di Siena and INFN, I-53100 Siena, Italy}
}

\shorttitle{D. Tescaro \etal The readout system of the MAGIC-II Cherenkov Telescope}
\maketitle

\begin{abstract}
In this contribution we describe the hardware, firmware and software components of the readout system of the MAGIC-II Cherenkov telescope 
on the Canary island La Palma.

The PMT analog signals are transmitted by means of optical fibers from the MAGIC-II camera to the 80 m away counting house where they are routed to the new high bandwidth and fully programmable receiver boards (MONSTER), which convert back the signals from optical to electrical ones. 
Then the signals are split, one half provide the input signals for the level ONE trigger system while the other half is sent to the digitizing units. 
The fast Cherenkov pulses are sampled by low-power Domino Ring Sampler chips (DRS2) and temporarily stored in an array of 1024 capacitors. 
Signals are sampled at the ultra-fast speed of 2 GSample/s, which allows a very precise measurement of the signal arrival times in all pixels. 
They are then digitized with 12-bit resolution by an external ADC readout at 40 MHz speed. 
The Domino samplers are integrated in the newly designed mezzanines which equip a set of fourteen multi-purpose PULSAR boards. 
Finally, the data are sent through an S-LINK optical interface to a single computer.

The entire DAQ hardware is controlled through a VME interface and steered by the slow control software program (MIR). 
The Data AcQuisition software program (DAQ) proceeds finally to the event building and data storage. 
\end{abstract}

 
\section{Introduction}
MAGIC-II is  the second of the two Cherenkov telescopes which constitute the MAGIC 
telescopes.
The array is located on the Canary Island of La Palma (2200~m above sea level, $28^\circ$45'N, $17^\circ$54'W).
The MAGIC telescopes are designed for the detection of VHE gamma rays with high sensitivity ($<$~1\% of the Crab Nebula flux is expected to be revealed in 50~h \cite{colin}) and the lowest energy threshold within the current generation of Cherenkov telescopes.\\
All the major components of MAGIC-II are already installed and the telescope is currently in its commissioning phase.
The first light ceremony took place on April 24-25th of this year.\\
The gamma ray signals are very short in time and therefore a very fast readout electronics is required.
The analog signals coming from the camera are very fast (the PMTs provide 2.5~ns wide pulses \cite{camera}).
The pulses have to be first processed to generate the trigger signal and then ultra-fast digitized.
Afterwards they have to be stored for the subsequent analysis \cite{abe}. 
A fast readout allows one to minimize the integration time and thus to reduce the influence of the background from the light of the night sky. 
With a fast sampling also the arrival time information can be obtained and exploited to improve the background discrimination \cite{timing}.\\
The single telescope acquisition rate is of the order of few hundred events per second, which requires to handle and store in the disk a data flow of the order of 100~MByte/s.\\
The readout system of the second MAGIC telescope differs significantly from the readout of MAGIC-I (see \cite{mux}).
New receivers boards have been designed (section \ref{sec:rec}).
The digitization of the analog signals is now performed through the so-called DRS2 chip \cite{drshome}.
A dedicated set of electronic boards have been developed for this purpose (section \ref{sec:domino}).
The slow control software, which guarantee the correct and stable functioning of the readout electronics, has been newly developed (section \ref{sec:mir}).
The final data acquisition software is a multi-thread C++ program based on the MAGIC-I DAQ and adapted to the new hardware features (section \ref{sec:daq}).\\

\section{Receiver board \label{sec:rec} } 
The receiver board interfaces the analog signals from the telescope camera with the level ONE trigger system and the digitization electronics.

%
 \begin{figure}[!t]
  \centering
  \includegraphics[width=2.5in]{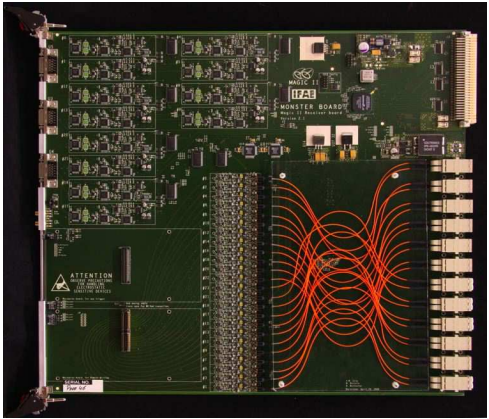}
  \caption{Picture of the MONSTER receiver board.}
  \label{rec_fig}
 \end{figure}

\noindent
A photo of the MONSTER (Magic Optical NanoSecond Trigger and Event Receiver) board is provided in figure~\ref{rec_fig}.
The board layout can be schematized in three parts: {\em analog}, {\em trigger} and {\em control}.\\
The optical fibers from the camera are connected to the analog channels of the receiver through LX5 connectors located on the back side of the board.
A pigtail routes the signal to a photodiode where it is converted back from optical to electrical analog signal.
The signal is then split in two branches, one which drives the signal to the digitization input and the second where the signal is shaped for the level ONE trigger.
An analog switch allows to inject in the analog channel an adjustable amplitude DC signal instead of the camera pulses. 
This is an useful tool for the calibration of the system.  

  \begin{table}[!h]
  \caption{Analog channel specifications.}
  \label{table_simple}
  \centering
  \begin{tabular}{|l|l|}
  \hline
    Total gain    &    18.5~dB  \\
Jitter              &             $<$100~ps\\
Bandwidth      &           750~MHz ($>$1~GHz capable) \\
Working range:       &  0.25$\div$1150~mV (differential output) \\ 
Inter-channel delay   & $<$1~ns\\
Crosstalk                  &   1:1000\\       
Noise (receiver only)                &   $<$200~$\mu$V RMS\\
Gain dispersion       &    30\% \\
  \hline
  \end{tabular}
  \end{table}
  \footnotetext{footnote inside table}

The trigger part represents the level ZERO of the trigger system chain.
If an analog pulse is above a certain threshold, a signal is generated that is sent in LVDS standard to the level ONE trigger.
This discriminator threshold is easily programmable for the individual channels so that a stable trigger rate can be guarantee even under variable light conditions (moonlight observations, stars in the field of view, etc.).
Thanks to dedicated channel rate counters the IPR (Individual Pixel Rate) information is online available.\\
The board also permits to inject test patterns in the trigger output for testing purposes.

 \begin{table}[!h]
  \caption{Level ZERO trigger specifications.}
  \label{table_simple}
  \centering
  \begin{tabular}{|l|l|}
  \hline
   Minimum input pulse                   &                  800~ps \\
   Work frequency  (pulse width 2.7ns)      &              $>$150~MHz  \\
   Digital programmable pulse width   &    2.7-12.7 ns  (10ps resolution)\\
   Digital programmable channel delay  &      0-20 ns  (10ps resolution)\\
   Signal transition time                        &               9.9 ns\\
   Jitter                                                &            $<$100~ps\\
    \hline
  \end{tabular}
  \end{table}

The firmware installed in the control part of the board is developed in VHDL code. 
It allows to monitor the board temperature, count the rate of the level ZERO trigger pulses up to the frequency of 50~MHz and manage all the control signals (trigger pattern, DC levels etc.).\\
The MONSTER is an highly integrated board: a single receiver includes 24 channels.
Maximum power consumption is of 75~W per board. 
The MONSTER board is fully compatible with the MAGIC-I telescope electronics.

\section{Digitization electronics \label{sec:domino} }

 \begin{figure*}[!t]
   \centerline{\subfloat[The PULSAR board]{\includegraphics[width=2.5in]{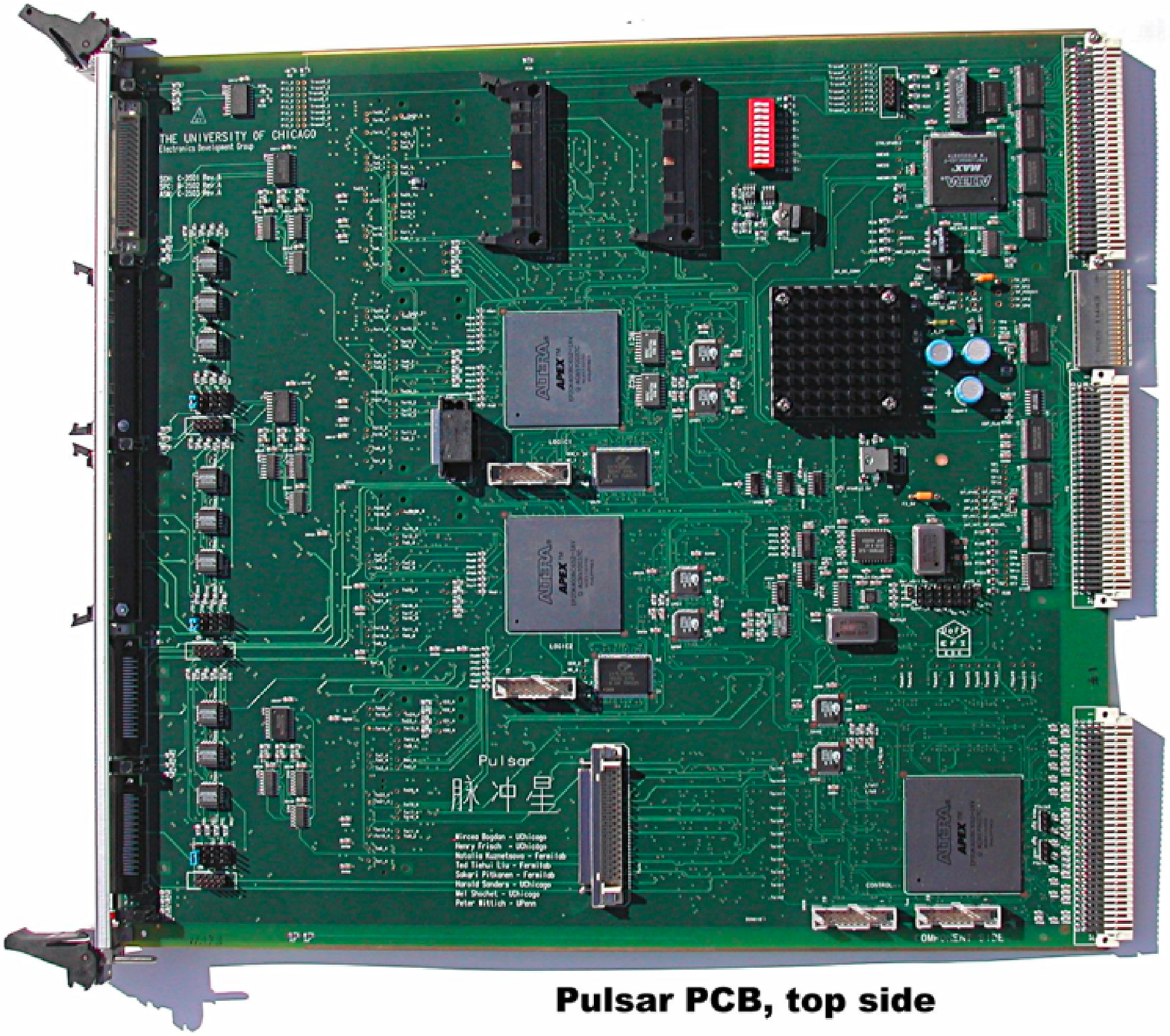} \label{sub_fig1}}
              \hfil
              \subfloat[The DRS2 chip and scheme of the mezzanine board]{\includegraphics[width=2.5in]{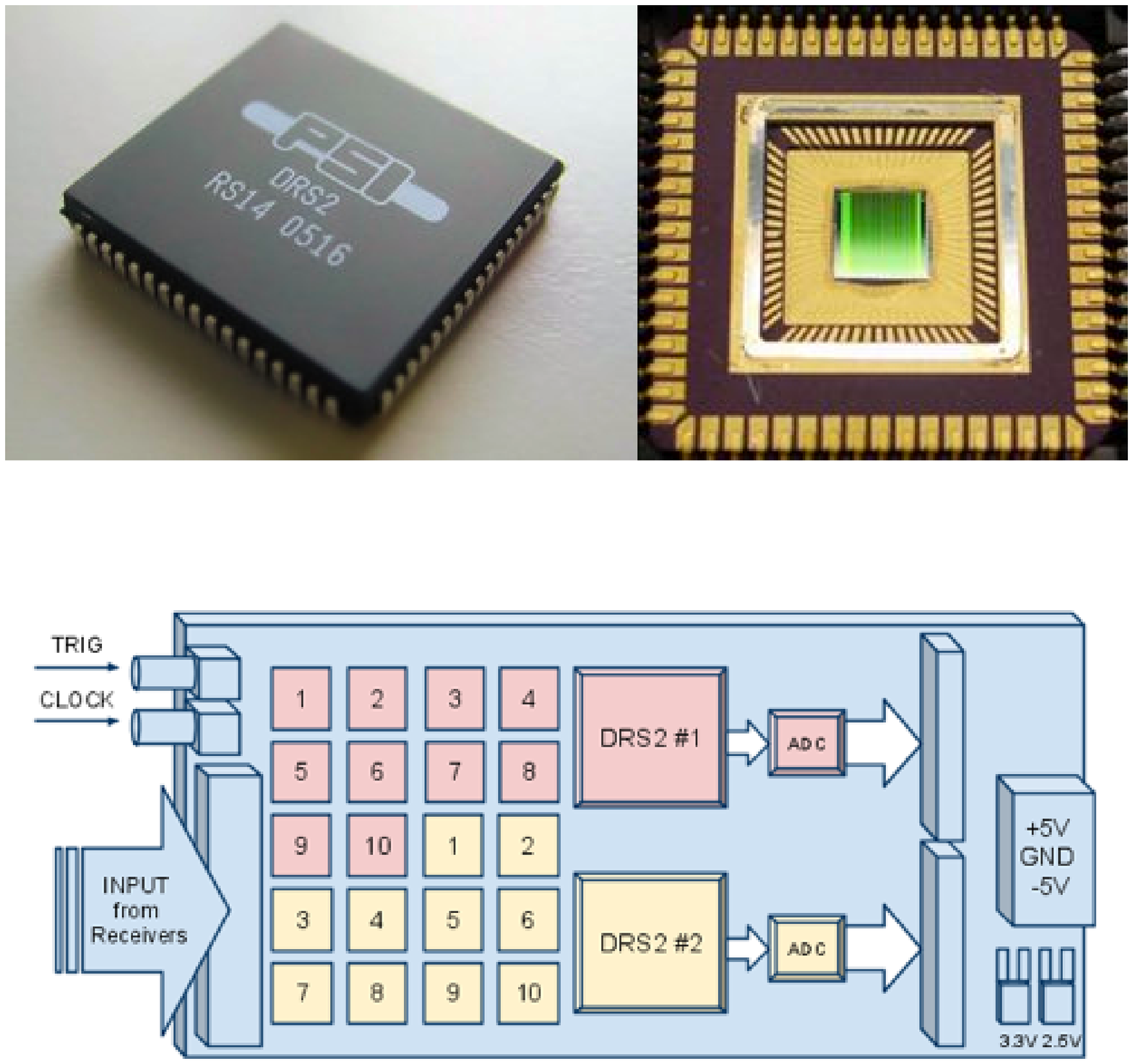} \label{sub_fig2}}
             }
   \caption{Picture of the PULSAR board (a) and scheme of the Domino mezzanine (b).}
   \label{dominos_fig}
 \end{figure*}

The digitization system for the MAGIC-II telescope \cite{bitossi} has been designed to fullfill the following  requirements:

\begin{itemize} 
    \item Ultra-fast sampling (2 GSample/s) of a large number of channels based on the DRS2 chip (Domino);
    \item Synchronous sampling of all channels to measure the timing structure of the camera images;
    \item Modularity to allow for  variations in  the number of input channels;
    \item Possibility to easily upgrade the sampling electronics from the DRS2 to the DRS-4 chip \cite{drshome};
    \item Flexibility and re-configurability to meet future changes on the operation of the experiment;
\end{itemize} 

Backbone of the readout is the PULSAR (PULSer And Recorder), a general purpose 9U~VME interface board for HEP applications developed by the University of Chicago \cite{pul}.
The general design philosophy of PULSAR is to use one type of motherboard (with a few powerful modern FPGAs and SRAMs) to interface any user data with any industrial standard link 
through the use of custom mezzanine cards. 
The picture \ref{dominos_fig}-(a) shows a PULSAR board with the main FPGAs: from top DATAIO-1, DATAIO-2 and CONTROL.
The DATAIO FPGAs handle the Dominos readout sequence and store the data into local SRAM memories. 
The CONTROL FPGA builds a formatted data packet to be transmitted to the DAQ computer.
The PULSAR board is used as a motherboard that hosts the mezzanines carrying the Domino analog samplers.

The mezzanine holding the Domino chips is a PMC card.
Four of these mezzanines are plugged on the rear side of the PULSAR board for a total of 80 channels each.
On the front panel are located the trigger and clock inputs, coming from the trigger and clock fan-out boards. 
These signals come via coaxial cables equipped with SMA connectors. 
The clock input is a 2~MHz signal that is used to synchronize the Domino PLL, ensuring in this way the same frequency and phase for all the samplers in the readout system.
On the bottom of figure \ref{dominos_fig}-(b) a sketch of the mezzanine is shown. 
A mezzanine is equipped with two DRS2 chips capable to simultaneously digitize 10 input channels each.
The Domino mezzanine has a +/-~5~V independent power supply. 
Secondary 3.3~V and 2.5~V are generated on each mezzanine with linear regulators.\\
The ADC has a nominal 12-bit resolution, the bandwidth of the DRS2 is 200~MHz. 

A special PULSAR board, named {\em digital}, is specially programmed to sample the trigger signal and to measure its arrival time.
This information is transmitted to the rest of the PULSAR boards, named {\em analog}, for reading out the region of interest only.
External digital information, like time stamp or trigger information, can be introduced in the data-flow via two front panel connectors.
Each connector carries 32 bits of digital information for a total of 64 bits per PULSAR.
Another dedicated PULSAR ({\em busy}) provides the busy signal that stops triggers when the readout system is busy processing one event.

The full readout system is made by using 14 analog PULSAR plus 2 special PULSAR boards, hosted in two 9U VME crates. 
The currently installed system is capable to readout a total of 1132 channels. 


\section{Slow Control program \label{sec:mir} }
The slow control of the MAGIC-II readout system is called
\textit{Magic Integrated Readout} (MIR). 
It is a multithread C-program, using which the following subsystems can be configured and monitored:
\begin{itemize}
  \item Trigger system (level ONE, level TWO and level THREE also known as {\em Stereo Trigger});
  \item PULSAR boards (including Dominos);
  \item Receiver boards;
  \item Telescope Calibration Unit;
\end{itemize}
The communication between the program and the hardware components is managed through the VME bus.  The hardware components controlled by MIR (in particular PULSAR boards, receiver boards, Telescope Calibration unit and the stereo trigger) are installed in seven WIENER VME crates.  
A standard PC, where MIR is installed, is equipped with a CAEN 32-bit 33 MHz PCI card A2818.  
This module allows control of the CONet: the network of up to eight daisy chained VME-PCI bridges.  
Each VME crate is equipped with VME-PCI Optical Link Bridge modules CAEN V2718, allowing for an optical fiber based daisy chain control of all crates.
Through MIR a communication with any component inside the seven crates via the VME optical chain is possible by accessing a given physical address of the component.

One of the main features of MIR is that it integrates all components of the readout (but the DAQ itself, see below) allowing for a close interaction between individual components needed during data taking as well as for the Domino calibration procedure.
MIR can be controlled via a command shell (local mode) or, alternatively, remotely through the Central Control program \cite{roby} (remote control). 
The tasks of the MIR are:
\begin{itemize}
  \item Configuring and monitoring level ONE trigger;
  \item Configuring and monitoring receivers:
    \begin{itemize}
      \item Setting pixel descriminator thresholds
      \item Setting pixel trigger delays 
      \item Setting width of pixel trigger signals 
      \item Individual pixel rate control
    \end{itemize}
  \item Configuring and monitoring PULSAR boards:
    \begin{itemize}
      \item Configuring domino settings
      \item Calibration of dominos
      \item Configuration of the busy signal 
      \item Configuration of the global trigger system 
    \end{itemize}
  \item Configuring and monitoring the level TWO trigger:
    \begin{itemize}
      \item Configuration of the prescaler, selection of trigger and prescale factors
      \item Monitor of the level ONE trigger macrocell rates
      \item Monitor of the individual trigger rates
     \end{itemize}
  \item Configuring and monitoring the Telescope Calibration Unit:
    \begin{itemize}
      \item Pulse injection triggers
      \item Calibration triggers
      \item Pedestal (random) triggers
      \item Setup of the calibration box, which includes calibration laser and filter wheels
    \end{itemize}
  \item Configuration of the global trigger system:
    \begin{itemize}
      \item Configuration of the global trigger logic
      \item Configuration of the trigger signal widths
      \item Automatic delay adjustment of trigger signals with telescope orientation
      \item Measurement of the trigger arrival time distribution
    \end{itemize}
\end{itemize}

\section{DAQ program \label{sec:daq} }
The Data AcQuisition (DAQ) program for MAGIC-II is named \emph{DominoReadout}. 
It is a multithread C++ program, based on \emph{MuxReadout}, the MAGIC-I data acquisition program.
The tasks of this software can be summarized as follow:
\begin{itemize} 
	\item Read the data samples from the readout hardware;
	\item Perform the event building merging the data from the different channels;
	\item Ensure the integrity of the events and detect data corruption;
	\item Perform an online analysis of the data (including a calibration of the DRS2 chip response);
	\item Store the data as raw files in a dedicated raid disk system.
\end{itemize}

  \begin{figure}[!t]
  \centering
  \includegraphics[width=2.5in]{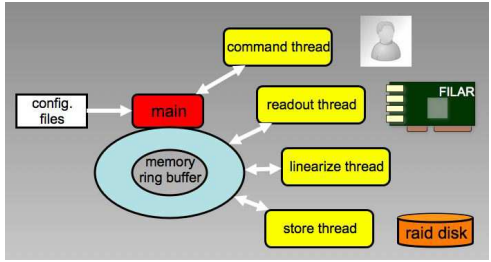}
  \caption{Scheme of the DAQ program.}
  \label{daqscheme}
 \end{figure}

\noindent
Each of these tasks is done in parallel thanks to the multithread 
 architecture of the program.
A scheme of the DAQ structure is presented if figure \ref{daqscheme}. 
Configuration text files are used to steer the DAQ initial parameters.
Commands can be given either from computer console or remotely from the Central Control program.\\
To easily handle the data in parallel the events are temporary copied on a ring buffer structure 4000 events deep, accessible synchronously by the running processes.
The more important elements are the {\em reading}, the {\em analyzing} and the {\em writing} threads.\\
The {\em Reading Thread} is appointed to read the data samples from the front-end  electronics of the readout.
The DAQ program interfaces the hardware through the FILAR board and its drivers \cite{slink}. 
The FILARs are PCI cards installed in the DAQ computer. 
This card, together with the HOLA board, constitutes the S-LINK optical data transfer link \cite{slink} (developed for the LHC experiment at Cern). 
The FILAR cards share part of the computer RAM memory with operative system of the PC. 
This memory buffer is used as a bridge to exchange the data within the hardware and the DAQ.
The reading thread constantly checks if new data are available in the shared memory area and if it is the case proceeds to the event building.
The event is then copied in a free segment of the ring buffer.
Integrity checks to avoid data corruption are performed at this stage. \\
The {\em Analyzing Thread} is appointed to two main tasks:
The calibration of the non perfect linear response of the DRS2 chip and the online analysis of the data.
The first task, called "domino calibration", corrects the original ADC values using previously computed calibration curves (specific for each capacitor of the DRS2 chips).
Since this process involves the manipulation of all the samples of an event (typically 1039 pixels with 80 samples each) the computer overhead due to this task is very high. 
In order to exploit the full potential of the machine, several copies of this thread are operated in parallel.   
The online analysis consists in a simple signal intensity and arrival time reconstruction which allows to online display the shower images on the Central Control monitor.\\
The {\em Writing Thread} removes the already analyzed events from the ring buffer and finally stores them on a raid disk (a 16~TByte SATA disks cabinet).
The MAGIC-II raw data files have a typical size of 2~GByte.
The data volume nightly generated might exceed 1~TByte.\\
The DAQ program is installed on a powerful 4-CPU 3.0~GHz server with 4~GByte RAM memory running Scientific Linux.
A sustainable trigger rate of 1kHz can be achieved (limited to $\sim$300~Hz if the online domino calibration is performed) corresponding to a 200~MByte/s data storage rate.


\section{Acknowledgement}
We would like to thank the 
Instituto de Astrofisica de Canarias 
for the excellent working conditions at the 
Observatorio del Roque de los Muchachos in La Palma. 
The support of the German BMBF and MPG, the Italian INFN 
and Spanish MCINN is gratefully acknowledged. 
This work was also supported by ETH Research Grant 
TH 34/043, by the Polish MNiSzW Grant N N203 390834, 
and by the YIP of the Helmholtz Gemeinschaft.

The authors would also like to thank Altera Inc. for supporting the project through the Altera University Program.
%
 


\begin{thebibliography}{99}
	\bibitem{colin} P.~Colin et al. \emph{Performance of the MAGIC telescopes in stereoscopic mode}, these proceedings.
 	\bibitem{camera} D.~Borla~Tridon et al. \emph{Performances of the Camera of MAGIC II Telescope}, these proceedings.
	\bibitem{abe} A. Moralejo et al. \emph{MARS, the MAGIC Analysis and Reconstruction Software}, these proceedings.
	\bibitem{timing} E.~Aliu et al.,
	Astroparticle Physics, 2009, vol. 30, p. 293-305.
	\bibitem{mux} F.~Goebel et al. \emph{Upgrade of the MAGIC Telescope with a Multiplexed Fiber-Optic 2GSamples/s FADC Data Acquisition System system}, XXX International Cosmic Ray Conference, 2008, vol. 3, p.1481-1484.
	\bibitem{drshome} \url{http://drs.web.psi.ch}.
	\bibitem{bitossi} M. Bitossi \emph{Ultra-Fast Sampling and Readout for the MAGIC-II Telescope Data Acquisition System }, PhD Thesis, 2009, \url{http://wwwmagic.mppmu.mpg.de/publications/theses/index.html}.
	\bibitem{pul} \url{http://hep.uchicago.edu/~thliu/projects/Pulsar/}.
	\bibitem{roby} R. Zanin and J. Cortina \emph{The Central Control of the MAGIC telescopes}, these proceedings.
	\bibitem{slink} \url{http://hsi.web.cern.ch/hsi/s-link/}.
  \end{thebibliography}
\end{document}